\documentclass[apaper,4pt]{article}
\usepackage{caption}
\usepackage{multicol}
\usepackage{graphicx}
\usepackage[left=3cm, right=3cm, bottom=1cm, top=4.2cm]{geometry}
\usepackage{amsmath}
\usepackage{amssymb}
\usepackage{setspace}
\usepackage{array}
\usepackage{subfigure}

\begin{document}

\begin{center}
{\bf \Large {Particle Ratios From Strongly Interacting Hadronic Matter   } }\\
\vspace{5mm}

\vspace{2mm}
{\small  Waseem Bashir, Saeed Uddin, Hamid Nanda}\\
\vspace{3mm}
{ \bf Department of Physics}\\
{\bf Jamia Millia Islamia (A Central University), New Delhi,
 India}
\end{center}
\doublespacing
\begin{abstract}
We calculate the particle ratios $K^+/\pi^+, \, K^-/\pi^-, \,\Lambda/\pi^-$ for a strongly interacting hadronic matter using non-linear Walecka model (NLWM) in relativistic mean-field (RMF) approximation. It is found that interactions among hadrons modify $K^+/\pi^+, \, \Lambda/\pi^-$ particle ratio while as $K^-/\pi^-$ is found to be insensitive to these interactions.    
\end{abstract}
\vspace{7mm}

\noindent {\bf I. Introduction}

Since the discovery of asymptotic freedom \cite{wil1973} in case of non-abelian gauge field theories, it was postulated that a phase transition from nuclear state of matter to quark matter is possible. It was further argued that this phase transition can take place at sufficiently high temperature and /or densities and can result in the transformations of hadrons into a new state of matter dubbed as quark-gluon plasma (QGP). Since then a considerable effort has been put forward to create and understand the properties of this new state of matter (QGP) and the corresponding phase transition. In order to study the dynamics of any phase transition in general a complete description of a given state of matter on the basis of some underlying theory is required. To understand the dynamics of quark-hadron phase transition the equation of state for both QGP phase and the hadronic phase is required. The QGP phase so far has been fairly described using Lattice Gauge theory, in case of vanishing or low baryon chemical potential. However the description of strongly interacting hadronic phase in terms of fundamental theory of strong interactions has proved to be far from trivial. This is primarily due to strong coupling among hadrons due to which the conventional methods of quantum field theory for example perturbative analysis does not remain valid for the description of such strongly interacting hadronic phase. Therefore one has to rely on alternate methods to describe the properties of hadronic phase  for example hadron resonance gas models, chiral models, quasi-particle models etc.    

However one can use another approach to determine the dynamics of strongly interacting hadronic phase and consequently of quark-hadron phase transition. By studying the spectra of hadrons one can in principle comment on some of the properties of the strongly interacting hadronic matter. For example by studying $\bar p/p$ ratio it has been argued that transparency effects in case of High energy Heavy-Ion collisions may become operative \cite{inam2016}. Recently  It has  been found that particle ratios of some of the hadrons  for example $K^+/\pi^+, K^- /\pi^-$ and $\Lambda/ \pi^-$ show a sudden rise for a specified range of center of mass energy $\sqrt {S_{NN}}$ in case of Heavy-Ion collisions \cite{NA49}. Taking into account the dependence of baryon chemical potential $\mu_B$ and temperature $T$ on the variable $\sqrt {S_{NN}}$ one can infer that the behaviour of these particle ratios may be sensitive to the critical region of quark-hadron phase transition. In this article we therefore evaluate the particle ratios $K^+/\pi^+, K^- /\pi^-$ and $\Lambda/ \pi^-$ for a strongly interacting hadronic matter and analyse their behaviour near first order quark-hadron phase transition. For hadronic phase we use non-linear Walecka model within relativistic mean-field (RMF) approximation. RMF theory has been  widely and successfully used to describe the properties of the nuclear matter and finite nuclei. Further RMF theory has been also used to describe the equation of state for strongly interacting dense hadronic matter for the application in supernova and neutron stars \cite{serot1986, hirata1996, Ren2002,shen2006, glad19855, zim1990, del1995,shen1998,shen2002}. In  RMF theory hadrons interact via the exchange of scalar and vector mesons and the interaction strength or coupling among hadrons is determined by different methods. For example the nucleon-meson coupling constants are determined by reproducing the ground state properties of the finite nuclei or by using nuclear matter properties that is discussed in Section-II. To describe the quark-gluon plasma (QGP) phase we use a Bag model equation of state.


\noindent{\bf II.	Model}\\ 
{\bf A. Hadronic Phase: Baryons }

The equation of state for asymmetric baryonic matter is presented in this section. To describe baryonic matter we use relativistic non-linear Walecka model (NLWM). In this model the interaction between baryons is governed by the exchange of various mesons. We include in this model following baryons $(p, n, \Sigma^+, \Sigma^0, \Sigma^-, \Xi^-, \Xi^0, \Lambda)$ along with their anti-particles. The interaction between baryons is carried by the exchange of neutral $\sigma$, isoscalar-vector $\omega$, isovector-vector $\rho$ and two additional hidden strangeness mesons $\sigma^*$ and $\phi$. In this model the Lagrangian density for baryons reads 
{\normalsize \begin{align*}
\ \mathcal L_{Bary}=& \,\sum_{\alpha} {{\overline\psi_{\alpha}}}\,[\gamma_{\mu} (i\partial^{\mu}  -g_{\omega {\alpha}}\omega^\mu-g_{\phi {\alpha}} \phi^\mu -
g_{\rho {\alpha}} {\vec{\tau}_{\alpha}}\cdot {{{\vec\rho}^{\, \mu}}})-M_{\alpha}^*] \psi_{\alpha} \\&+
\frac{1}{2}\partial_\mu\sigma \,\partial^\mu\sigma-U(\sigma)-
\frac{1}{4}W_{\mu\nu}W^{\mu\nu}+\frac{1}{2}m_\omega^2\omega_\mu\omega^\mu + \frac{1}{4}c_3(\omega_\mu\omega^\mu)^2 \\&-
 \frac{1}{4}{\vec R_{\mu\nu}} \cdot {\vec R^{\mu\nu}}+\frac{1}{2}m_\rho^2 \vec\rho_{\, \mu}\cdot \vec\rho^{\, \mu}+ \frac{1}{2}\partial_\mu\sigma^* \partial^\mu
\sigma^*-\frac{1}{2}m_{\sigma^*}^2 {\sigma^*}^2\\&-\frac{1}{4}S_{\mu\nu}S^{\mu\nu}+\frac{1}{2}m_\phi^2\phi_\mu\phi^\mu, \tag{1}
\end{align*}  
where $\Psi_{\alpha}$ is the fermionic field corresponding to baryon $\alpha$. The interaction between baryons is carried by the exchange of neutral $\sigma$, isoscalar-vector $\omega$, isovector-vector $\rho$ and two additional hidden strangeness mesons $\sigma^*$ and $\phi$. $U(\sigma)= \frac{1}{2} m_{\sigma}^2 \sigma^2 + \frac{1}{2} g_2 \sigma^3 +\frac{1}{2} g_3 \sigma^4$ is the scalar self interaction term for $\sigma$ field. Also $S^{\mu \nu} = \partial^{\mu} \phi^{\nu}- \partial^{\nu} \phi^{\mu}$, $\vec R^{ \mu \nu}=\partial^{\mu} \vec\rho^{\,\nu}-\partial^{\nu} \vec\rho^{\,\mu}\,+ \,g_{\rho}\left(\vec\rho^{\, \mu} \times \vec\rho^{\, \nu} \right),W^{\mu \nu}= \partial^{\mu}\omega^{\nu}-\partial^{\nu}\omega^{\mu}$ and $\frac{1}{4} c_3 \left(\omega_{\mu}\omega^{\mu} \right)^2$ is the Bodmer correction or self-interaction term for the vector field $\omega^{\mu}$ and $g_{i \alpha}$ are the coupling constants that characterise the strength of interaction between mesons $i= \omega, \sigma, \sigma^*, \rho, \phi$ and baryons $\alpha= p,n, \Lambda, \Sigma, \Xi$. Here $M^*_{\alpha}= M_{\alpha} +g_{\sigma \alpha} \sigma + g_{\sigma^* \alpha} \sigma^*$ is in-medium mass of baryon, where  $M_{\alpha}$ is the bare-mass of baryon.  Also $m_i$ is the mass of exchange mesons and $\vec \tau$ is the isospin operator. Using relativistic mean-field (RMF) approximation under which the field variables are replaced by their space-time independent classical expectation values that is  $ \sigma \rightarrow <\sigma>  = \sigma_0$,\,\, $\omega_{\mu} \rightarrow <\omega_{\mu}> = \delta_{\mu 0}\,\omega_0 $, \,$\vec \rho_{\mu} \rightarrow$ $<\vec \rho_{\mu}> = \delta_{\mu 0} \,\delta^{i3}\rho_{03}$ and $\phi_{\mu} \rightarrow <\phi_{\mu}>=  \delta_{\mu 0} \phi_{0}$ the thermodynamic potential per unit volume corresponding to Lagrangian density (1) can be written as
\begin{align*}
 \left(\frac{\Omega}{V}\right)_{Bary}\,\,= \,\,\,\,&\frac{1}{2}m_{\sigma}^2 \sigma_0^2 + \frac{1}{3} g_2 \sigma_0^3 + \frac{1}{4} g_3 \sigma_0^4- \frac{1}{2} m_{\omega}^2 \omega_0^2- \frac{1}{4}c_3 \omega_0^4- \frac{1}{2} m_{\rho}^2 \rho_{03}^2+ \frac{1}{2}m_{\sigma^*}^2 {\sigma_0^*}^2
-\frac{1}{2} m_{\phi}^2 \phi_0^2\\&-2T\sum_{\alpha}\int \frac{d^3 k}{(2 \pi)^3}\left\{\ln \left[1+ e^{-\beta\left( E_{\alpha}^*- \nu_{\alpha} \right)}\right] + \ln \left[1+ e^{-\beta( E_{\alpha}^*+ \nu_{\alpha})} \right]  \right\},  \tag{2}
 \end{align*}
 where effective baryon energy is $E_{\alpha}^* = \left(k_{\alpha}^2 + {M_{\alpha}^*}^2\right)^{{1}/{2}}$ and effective baryon chemical potential is $\nu_{\alpha}= \mu_{\alpha}-g_{\omega {\alpha}}\omega_0 -g_{\phi {\alpha}}\phi_0-g_{\rho {\alpha}}\tau_{3 {\alpha} }\rho_{03}$. Also parameter $\beta$ is $\beta= 1/T$, where $T$ is the temperature.\\
 
\noindent{\bf B. Hadronic Phase: Bosons (pions + Kaons) } 

 To incorporate bosons (pions + Kaons) in our model we use an approach similar to the one used to model baryonic phase that is  we use a meson exchange type of Lagrangian for bosons as well. The Lagrangian density in a minimal-coupling scheme is \cite{prak1994, glad1999}
 \begin{align*}
\mathcal L_{Bosons} = \sum_{b}D_{\mu}^* \Phi_{b}^* \,D^{\mu} \Phi_{b} - {m_b^*}^2 \Phi_{b}^* \Phi_{b},     \tag{3}
\end{align*}
 where $\Phi_b$ is the bosonic field with summation carried over bosons $b$. Here covariant derivative is
  \begin{align*}
D_{\mu}=& \,\partial_{\mu} + i X_{\mu},  \tag{4}  
\end{align*}
with the four vector $X_{\mu}$ defined as 
\begin{align*}
X_{\mu}\equiv& \, g_{\omega b}\, \omega_{\mu}\,+\,g_{\rho b}\, {\vec \tau_b} \cdot {\vec \rho_{\mu}} + g_{\phi b} \phi_{\mu}, \tag{5}
\end{align*}
and $m_b^*= m_b +g_{\sigma b} \sigma + g_{\sigma^* b} \sigma^*$ is the effective-mass of bosons. Also $g_{ib}$ are the coupling constants that characterise the strength of interaction between exchange mesons $i= \sigma, \omega, \rho, \phi, \sigma^*$ and bosons (pions + kaons). Here $\vec \tau_b$ is the isospin operator with its third component defined as
\begin{align*}
\tau_{3\pi} = +1 (\pi^+),\hspace{1mm} 0 (\pi^0),\hspace{1mm} -1 (\pi^-); \hspace{5mm} \tau_{3K}= +1/2 (K^+,  K^0), \hspace{1mm} -1/2 (K^-, \bar K^0). \tag{6}
\end{align*}

It has to be mentioned  that one can use even chiral perturbation theory \cite{kap1989} to describe bosons in the hadronic matter. In an earlier work  \cite{ellis1995} kaons were incorporated using chiral perturbation theory whereas baryons were incorporated using Walecka model. However in \cite{ellis2} it was put forward that this approach of modelling baryonic phase with Walecka model and bosonic phase with Chiral Lagrangian has some inconsistency that may influence the final results. In our approach baryons and bosons are incorporated using similar methodology that is using meson-exchange type Lagrangian and therefore this approach is expected to be more consistent. In RMF approximation, the thermodynamic potential for Lagrangian density (3) can be written as
\begin{align*}
\left(\frac{\Omega}{V}\right)_{Bosons}= \hspace{3mm} T\sum_{b} \Gamma_b \int \frac{d^3 k}{(2 \pi)^3} \left\{\ln [1-e^{-\beta({\omega_{b+}} -\mu_b)} ]+ \ln [1-e^{-\beta(\omega_{b-}+\mu_b)}]  \right\},  \tag{7}
\end{align*}
where $\omega_{b\pm}= E_b^* \pm X_0$, with effective energy defined as $E_b^* = \sqrt {k^2 +{m_b^*}^2}$ and $X_0$ is the temporal component of four-vector $X_{\mu} \equiv g_{\omega b} \omega_{\mu} +g_{\rho b} \vec \tau_b \cdot \vec \rho_{\mu} + g_{\phi b} \phi_{\mu}$. Also $\mu_b$ is the boson chemical potential and $\Gamma_b$ is the spin-isospin degeneracy factor of boson $b$.\\ 


\noindent {\bf C. Hadronic Phase: Field Equations }

The thermodynamic potential per unit volume for hadronic medium $(\Omega/ V)_H$ can be therefore written as
\begin{align*}
(\Omega/V)_H = (\Omega/V) _{Bary} + (\Omega/V)_{Bosons},  \tag{8}
\end{align*}
where $(\Omega /V)_{Bary}$ and $(\Omega /V)_{Bosons}$ are as defined in (2) and (7) respectively. 

The different thermodynamic observables of the hadronic system  for example entropy, pressure, number density  can be evaluated as follows
\begin{align*}
S(T,V,\mu_H) = -\frac{\partial \Omega_H}{\partial T} \bigg\arrowvert_{V,\mu_H}  \tag{9}\\
P(T,V,\mu_H) = - \frac{\partial \Omega_H}{\partial V} \bigg\arrowvert_{T, \mu_H}  \tag{10}\\
N(T,V,\mu_H) = - \frac{\partial \Omega_H}{\partial \mu_H} \bigg\arrowvert_{T,V},   \tag{11}
\end{align*}
 provided the expectation values of the exchange-mesons field variables $(\sigma_0, \sigma_0^*, \omega_0, \rho_0, \phi_0 )$ are known. 

To evaluate the expectation value of exchange-meson field variables one can solve following set of coupled equations of motion for different field variables that are obtained after minimising the action integral $S = \int ({\mathcal L_{Bary}} +{\mathcal L_{bosons}})\, d^4x$ with respect to different exchange-meson field variables, that is
 
\begin{align*}
 m^2_{\sigma} \sigma_0=& \,-g_2\sigma_0^2-g_3 \sigma_0^3- \sum_{\alpha} g_{\sigma \alpha} \frac{\Gamma_{\alpha}}{(2 \pi)^3} \int \frac{d^3k}{\sqrt{(k^2 +{M_{\alpha}^*}^2)}} M_{\alpha}^* \,(F_{\alpha}^{(+)}+ F_{\alpha}^{(-)}) - \frac{1}{2}\sum_b m_b\, g_{\sigma b} \frac{\Gamma_b}{(2 \pi)^3} \int {d^3 k}\\& \times  (\omega_{b+}^{-1} F_{b}^{(+)} + \omega_{b-}^{-1}F_{b}^{(-)}),  \tag{12}
 \end{align*}
 \begin{align*}
 m_{\omega}^2 \omega_0=&\,-c_3 \omega_0^3 + \sum_{\alpha} g_{\omega \alpha}\, n_{\alpha}  +\,\frac{1}{2}\sum_b g_{\omega b} \, n_b -\frac{1}{2}\sum_b g_{\omega b} X_0\frac{\Gamma_b}{(2 \pi)^3} \int {d^3k}\,\, (\omega_{b+}^{-1}F_{b}^{(+)} + \omega_{b-}^{-1} F_{b}^{(-)}), \tag{13}\\\\
  m_{\rho}^2 \rho_0 =& \sum_{\alpha} g_{\rho {\alpha}} \tau_{3\alpha} n_{\alpha} +\frac{1}{2} \sum_b g_{\rho b} \tau_{3b}\, n_b  -\frac{1}{2}\sum_b g_{\rho b}\tau_{3b} X_0 \frac{\Gamma_b}{(2 \pi)^3} \int {d^3k} \,\, (\omega_{b+}^{-1} F_{b}^{(+)} + \omega_{b-}^{-1}F_{b}^{(-)}),       \tag{14}
 \end{align*}
 \begin{align*} 
m_{\sigma*}^2 \sigma_0^*=& \sum_{\alpha} g_{\sigma^*{\alpha}} \frac{\Gamma_{\alpha}}{(2 \pi)^3} \int \frac{d^3k}{\sqrt{(k^2+ {M_{\alpha}^*}^2)}} M_{\alpha}^*\,(F_{\alpha}^{(+)}+F_{\alpha}^{(-)}) -\frac{1}{2}\sum_b m_b\, g_{\sigma^* b}\,  \frac{\Gamma_b}{(2 \pi)^3} \int {d^3k}\, (\omega_{b+}^{-1} F_{b}^{(+)} + \omega_{b-}^{-1}F_{b}^{(-)}), \tag{15}\\
 m_{\phi}^2 \phi_0 =& \sum_{\alpha} g_{\phi \alpha}\, n_{\alpha} + \frac{1}{2}\sum_b g_{\phi b} \,\,n_b - \frac{1}{2}\sum_b g_{\phi b} \frac{\Gamma_b}{(2 \pi)^3} \int {d^3k}\,\,(\omega_{b+}^{-1} F_{b}^{(+)} + \omega_{b-}^{-1}F_{b}^{(-)}),   \tag{16}
 \end{align*}\\  
where distribution functions for baryons $F_{\alpha}^{(+)}$ and anti-baryons $F_{\alpha}^{(-)}$ are given by
\begin{align*}
F_{\alpha}^{(\pm)} = \frac{1}{e^{\beta\left(E_{\alpha}^* \,\mp \, \nu_{\alpha} \right)}+1}, \tag{17}
\end{align*}
and net-baryon density is 
\begin{align*}
 n_{Bary} = \sum_{\alpha} n_{\alpha}=& \hspace{3mm} \sum_{\alpha} \frac{2}{(2 \pi)^3} \int (F_{\alpha}^{(+)}-F_{\alpha}^{(-)})\,  d^3k. \tag{18}
\end{align*}

Similarly the distribution function of bosons $F_b^{(+)}$ and their anti-particles $F_b^{(-)}$ is
 \begin{align*}
F_{b}^{(\pm)}= \frac{1}{e^{\beta(\omega_{b\pm}\mp \mu_b)}-1}= \frac{1}{e^{\beta[(\epsilon_b^* \pm X_0) \mp \mu_b]}-1} = \frac{1}{e^{\beta(\epsilon_b^* \mp \nu_b)}-1},  \tag{19}
\end{align*}
where $\nu_b= \mu_b -X_0$ is the effective chemical potential of boson $b$. Also the boson density is
 \begin{align*}
n_{bosons} = \sum_{b} n_{b} =\, \sum_b \Gamma_b \int \frac{d^3 k}{(2 \pi)^3} \left(F_{b}^{(+)}\,\,-F_{b}^{(-)} \right). \tag{20}
\end{align*}\\ 
 \noindent {\bf D. Hadronic Phase: The Coupling Constants }

To fix baryon-meson coupling constants we use two very successful parameter sets of RMF model namely parameter set TM1 and NL3. These parameter set are listed in Table-I. These parameters have been obtained by evaluating the ground state properties of finite nuclei \cite{sugahara1994,ring1997}. For meson-hyperon coupling constants we use quark model values of vector couplings. These are given by
$$ \frac{1}{3} g_{\omega N}= \frac{1}{2} g_{\omega \Lambda} = \frac{1}{2} g_{\omega \Sigma} = g_{\omega \Xi} ,$$ 
$$g_{\rho N} = \frac{1}{2} g_{\rho \Sigma} = g_{\rho \Xi},\,\,\, g_{\rho \Lambda} = 0,$$ $$2 g_{\phi \Lambda} = 2g_{\phi \Sigma} = g_{\phi \Xi} = - \frac{2 \sqrt 2}{3} g_{\omega N}, \,\,\, g_{\phi N}=0.$$

The potential depth for hyperons in baryonic matter is fixed as follows. Representing the potential depth of hyperon $h$ in baryonic matter $B$ as $U^{(B)}_h$ we use use $U_{\Lambda}^{(N)}= -28 $ MeV, $U_{\Sigma}^{(N)}= +30$ MeV and $U_{\Xi}^{(N)}= -18$ MeV to determine the value of scalar coupling constants $g_{\sigma \Lambda},\,\, g_{\sigma \Sigma}$ and $g_{\sigma \Xi}$ respectively \cite{gal1988,gal2000,bunta2004} . The hyperon couplings with strange mesons are restricted by the relation $U_{\Xi}^{(\Xi)} \simeq U_{\Lambda}^{(\Xi)} \simeq 2 \,U_{\Xi}^{(\Lambda)} \simeq 2\,\,U_{\Lambda}^{(\Lambda)}$ obtained in \cite{schaffner1994}. For the hyperon-hyperon interactions we use the square well potential with depth $U_{\Lambda}^{(\Lambda)}= -20$ MeV \cite{shaffner1993}. In Table-II we list the values of the coupling constants determined from these hyperon potentials. Next in Table-III and Table-IV  we give kaon-meson and pion-meson coupling constants that are used in our calculation. 

 Regarding  antibaryon-meson couplings \cite{mis2005} there is no reliable information particularly for high density matter. Therefore we will use in our work the values of the antibaryon-meson couplings that are derived using G-parity transformation. The G-parity transformation is analogous to ordinary parity transformation in configurational space which inverts the direction of three vectors. The G-parity transformation is defined as the combination of charge conjugation and rotation. The $\pi$ degree of rotation is done around the second axis of isospin space.

It is already known that exchange mesons $\sigma$ and $\rho$ have positive G-parity and $\omega $ and $\phi$ have negative G-parity. Therefore by applying G-parity transformation to nucleon potentials one can obtain the corresponding potential for anti-nucleons. The result of G-parity transformation can be written as $$g_{\sigma {\bar \alpha}} = g_{\sigma \alpha},\,\,\, g_{\omega {\bar \alpha}}= -g_{\omega \alpha},\, \,\,g_{\rho \bar{\alpha}}= g_{\rho \alpha},\,\,\,  g_{\phi {\bar \alpha}}= -g_{\phi \alpha},$$
where $\alpha $ and $\bar \alpha$ denote baryons and antibaryon's respectively.
It is worthwhile to mention here that kaon-meson couplings can be fixed for two different kaonic potentials namely strongly attractive potential UK[S] and weakly attractive potential UK[W]. These are given  in Table-III. Finally the pion-meson couplings  are given in Table-IV. Since the pion is non-strange particle therefore its coupling with strange mesons $\sigma^*$ and $\phi$ is essentailly zero.\\\\
{\bf III.\,\, Results and Discussions}

In this article our main aim is to evaluate the properties of strongly interacting hadronic matter at finite temperature. Therefore we make an attempt to analyse the properties of particle ratios $K^+/ \pi^+, \, K^-/\pi^-, \,\Lambda/\pi^-$ for this matter.   
Using the values of baryon-meson and boson-meson coupling constants as defined in the previous section, one can solve the coupled integral equations for the field variables $(\sigma_0, \omega_0, \rho_0, \sigma^*_0, \phi_0)$ and consequently one can obtain the thermodynamic observables of the hadronic system for a given values of temperature $T$ and baryon chemical potential $\mu_B$.  In the discussion to follow we use following parameter sets for fixing baryon-meson and boson-meson coupling constants used in this model. For baryon-meson couplings we use parameter set TMI as listed in Table-I and Table-II. For kaon-meson coupling constants we use parameter sets TM2 and GL85 that correspond to strongly attractive and weakly attractive kaonic potentials respectively and are listed in Table-III. The pion-meson coupling constants are listed in Table-IV. In the following analysis we will impose the strangeness conservation criteria $S- \bar S$ = 0, where $S$ and $\bar S$ are total strangeness and anti-strangeness of the system under consideration.

In Fig.(1) we plot the variation of $K^+/\pi^+$ with temperature for fixed values of baryon chemical potential. For $\mu_B = 0$MeV we show $K^+/\pi^+$ for hadronic matter with coupled baryons and bosons for two parameter sets TM1 and NL3 with kaonic potential UK[S] and UK[W] respectively. The result obtained for non-interacting hadrons is calculated using hadron resonance gas (HRG) model. The  effect of chiral symmetry is analysed by decoupling Nambu-Goldstone modes (pions and kaons) from baryons and result is denoted by RMF[D]. This is obtained by setting $g_{\sigma b} = g_{\sigma^* b} = g_{\omega b} = g_{\phi b} = g_{\rho b} = 0 $. It is worthwhile to mention here that with this choice of boson-meson couplings, baryons can still interact strongly via the exchange of mesons ($\sigma, \sigma^*, \omega, \rho, \phi$) while as bosons (pions + kaons) get decoupled from baryons and hence remain in the system as free particle with no interaction.   One can see that for $\mu_B = 0$MeV  the effect of interaction is neglible on $K^+/\pi^+$ ratio even upto relatively high temperature of about $T = 150$MeV. For higher baryon chemical potential that is $\mu_B = 450$MeV one can see that the interactions modify  $K^+/\pi^+$ ratio. For hadronic phase with coupled baryons and bosons the effect of interaction is to increase $K^+/\pi^+$ ratio. Further one can see that the rise of $K^+/\pi^+$ ratio is significant, only in the critical region CR of first order quark-hadron phase transition, that corresponds to large value of bag constant $B$. Here we have calculated critical region with equation of state for Quark Gluon Plasma that is consistent with Lattice QCD (see Appendix-1). The Bag value was fixed in the range $B^{1/4} = 165-200$ MeV.  However on decoupling baryons from bosons one can see that the ratio $K^+/\pi^+$ drops  and infact becomes slightly lower than HRG model result.  

 In Fig.(2) we plot the variation of $K^+/\pi^+$ with baryon chemical potential $\mu_B$ at fixed value of temperature $T= 50$MeV and $T =70$MeV. At lower tempearture $K^+/\pi^+$ ratio for NL3 parameter set is same as that for HRG model. While as for TM1 parameter set $K^+/\pi^+$ ratio is less as compared to that of HRG model. However at higher temperature the particle raio $K^+/\pi^+$ for both parameter sets that is  TM1 and NL3 is less as compared to HRG model.   
 
In Fig.(3) we next plot the variation of particle ratio $\Lambda/\pi^-$ with temperature for fixed values of baryon chemical potential $\mu_B$. At lower baryon chemical potential that is $\mu_B = 0$MeV the effect of interaction among hadrons is negligible on particle ratio $\Lambda/\pi^-$ even for relatively high temperature of $T = 150$MeV. However for higher baryon chemical potential that is $\mu_B = 450$MeV the effect of interaction among hadrons  increases the ratio $\Lambda/\pi^-$.  One can again see that the rise of $\Lambda/\pi^-$ is somewhat significant in the critical region of phase transition that corresponds to a large value of bag constant $B$. Here CR denotes the critical region of first order quark-hadron phase transition that is calculated with bag value $B^{1/4} = 165-200$MeV and lattice motivated equation of state (see Appendix-1). Interestingly one can see that the effect of interaction on  $\Lambda/\pi^-$ vanishes if bosons decouple from baryons. $\Lambda/\pi^-$ ratio for a system of strongly interacting hadrons with baryons decoupled from bosons is represented by RMF[D].         

In Fig.(4) we next plot the variation of particle ratio $\Lambda/\pi^-$ with baryon chemical potential at fixed values of temperature $T = 50$MeV and $T = 70$MeV. At lower temperature that is $T = 50$MeV the particle ratio $\Lambda/\pi^-$ behaves differently under parameter sets TM1 and NL3.  However at large temperature that is $T = 70$MeV the particle ratio $\Lambda/\pi^-$  for parameter sets TM1 and NL3 coincide with non-interacting systems even upto very large values of baryon chemical potential.

In Fig.(5) we next plot the variation of particle ratio $K^-/\pi^-$ as a function of temperature for fixed baryon chemical potential that is $\mu_B = 0$MeV and $\mu_B = 450$MeV. For lower value of baryon chemical potential that is $\mu_B = 0$MeV the effect of interactions among hadrons with coupled baryons and bosons is negligible on the particle ratio $K^-/\pi^-$.  Interestingly even for higher baryon chemical potential that is $\mu_B = 450$MeV the effect of interaction is still negligible on $K^-/\pi^-$ ratio. However on decoupling bosons from baryons the $\Lambda/\pi^-$ ratio becomes large at higher values of temperature. For a reference we have again shown the critical region of first order quark-hadron phase transition CR. Here we have used equation of state for quark-gluon plasma phase with Bag value $B^{1/4} = 165-200$MeV and strong coupling constant $\alpha_S = 0.2$.     

Finally in Fig.(6) we show the variation of particle ratio $K^-/\pi^-$ with baryon chemical potential $\mu_B$ for fixed values of temperature $T = 50$MeV and $T = 70$MeV. For the case of strongly interacting hadronic matter with coupled baryons and bosons  the particle ratio $K^-/\pi^-$ is same as that of a system of non-interacting hadrons, even upto very large baryon chemical potential values.															

In Fig.(7) we plot the variation of effective chemical potential $\nu_{Hadrons}$ and effective mass $(m^*)$  to bare mass $(m)$  ratio $m^*/m$ of hadrons as a function of temperature for fixed values of baryon chemical potential $\mu_B = 0$MeV and $\mu_B = 450$MeV. Next in Fig.(8) we have shown the variation of these observables that is $\nu_{Hadrons}$ and $m^*/m$ as a function of baryon chemical potential $\mu_B$ at fixed values of temperature $T = 50$MeV and $T = 70$MeV. These observables have been calculated for a system of strongly interacting hadronic system with coupled baryons and bosons, with parameter set TM1. For kaon-meson coupling we use parameter set TM2 that corresponds to strongly attractive kaonic potential UK[S].  To complete our discussion we next plot in Fig.(9) the variation of strange chemical potential $\mu_S$ as functions of baryon chemical potential $\mu_B$ and temperature $T$ at fixed values of temperature and baryon chemical potential, respectively.  The strangeness chemical potential is fixed by imposing the constraint of strangeness conservation $(S- \bar S = 0)$ in the hadronic medium. The values of strangeness chemical potential $\mu_S$ shows a decrease with increasing baryon chemical potential with increasing temperature. Further we also observe that the strangeness chemical potential decreases with the increase of temperature for increasing values of baryon chemical potential. This is consistent with the hadron resonance gas model \cite{soll}.\\\\
{\bf IV.\,\, Summary}

In this article we have calculated particle ratios $K^+/\pi^+,\,K^-/\pi^-,\, \Lambda/\pi^-$ for a strongly interacting hadronic matter using non-linear Walecka model in relativistic mean-field (RMF) approximation. In the hadronic medium we incorporate baryons and bosons (pions + Kaons). To describe baryons and bosons we use a meson-exchange type of Lagrangian and evaluate thermodynamic observables of hadronic matter in RMF approximation. It is found that the interaction among hadrons which in the present model is  mediated by the exchange of $\sigma, \sigma^*, \omega, \rho $ and $\phi$ mesons can result in the modification of $K^+/ \pi^+, \Lambda/ \pi^-$  ratio, while as the particle ratio $K^-/\pi^-$  is found to be independent of the interaction among hadrons.\\

\noindent {\bf Acknowledgements}\\
Waseem Bashir is greatful to Council of Scientific and Industrial Research, New Delhi for awarding Research Associateship.\\

\noindent{\bf V.\,\, APPENDIX-1}\\
\noindent {\bf  Quark-Gluon Plasma Phase:} In this section we present equation of state for quark-gluon plasma (QGP) phase. We consider three quark flavours up (u), down (d), Strange (s) and gluons. We use a Bag model equation of state with perturbative corrections of the order of $\alpha_s$ that is consistent with Lattice data \cite{satarovkarsch2001}. The pressure $P$ and energy density $\epsilon$ are given by

\begin{align*} 
P_{QGP} = & \left(\tilde N_g + \frac{21}{2} \tilde N_f \right)\frac{\pi^2 T^4}{90}+ \tilde N_f \left(\frac{\mu^2 T^2}{18}+ \frac{\mu^4}{324 \pi^2} \right)+
            \frac{1-{\zeta}}{\pi^2}\int_{m_{s}}^{\infty}dE \left(E^2 -m_{s}^2\right)^{\frac{3}{2}} \left(F_{q}^{(+)}+{{F}}_{q}^{(-)}\right)\\ &- B \tag{A1} 
\end{align*}
\begin{align*}
\varepsilon_{QGP}= & \left(\tilde N_g + \frac{21}{2} \tilde N_f \right)\frac{\pi^2 T^4}{30}+ 3\tilde N_f \left(\frac{\mu^2 T^2}{18}+ \frac{\mu^4}{324 \pi^2} \right)+3\left(
            \frac{1-{\zeta}}{\pi^2}\right)\int_{m_{s}}^{\infty}dE \left(E^2 -m_{s}^2\right)^{\frac{3}{2}} \left(F_{q}^{(+)}+{{F}}_{q}^{(-)}\right)\\ &+B, \tag{A2}
\end{align*}
where  ${\tilde N_{g}} = 16(1-\frac{4}{5}{\zeta})$ and ${\tilde  N_{f}}= 2(1-{\zeta})$ are the effective number of gluons and fermions respectively. The quark chemical potential is $\mu_q$ and $\zeta =\alpha_s$ is the coupling constant. For the determination of coupling constant $\alpha_s$ see for example \cite{alford}. Here $F_q^{(+)}$ and ${ F}_{q}^{(-)}$are the Fermi-Dirac distribution functions for quarks and antiquarks respectively and B is the Bag constant. To see the possible effect of Bag constant on various observables see for example \cite{thoma2000}. In the above calculation up (u) and down (d) quarks are considered massless while as strange quark (s) is of finite mass $m_s = 150$MeV. In the present discussion heavy quark flavours have not been considered.

\newpage
\newpage
\begin{figure}[htp]
\includegraphics[scale=1.2]{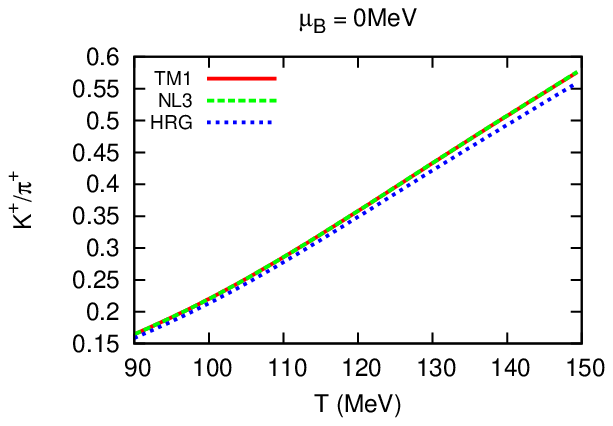}
\includegraphics[scale=1.2]{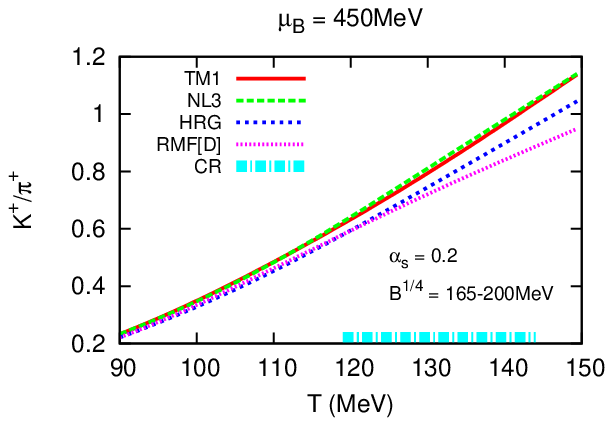}
\caption{{ $K^+/\pi^+$ ratio as a function of temperature $T$ for fixed value of baryon chemical potential $\mu_B = 0$MeV and $\mu_B = 450$MeV for (a). Strongly interacting hadronic matter with coupled baryons and bosons with two parameter sets TM1 and NL3 for kaonic potential UK[S] and UK[W] respectively, (b). Strongly interacting hadronic matter with decoupled baryons and bosons RMF[D] and (c). Hadron Resonance Gas (HRG) model. CR is the critical region of first order quark-hadron phase transition calculated with Bag value $B^{1/4} = 165-200 $MeV. $\alpha_S$ is the coupling strength for quarks.}}
\end{figure}

\begin{figure}[htp]
\includegraphics[scale=1.2]{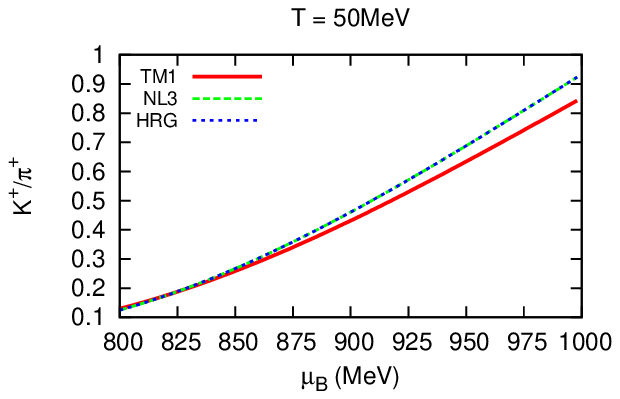}
\includegraphics[scale=1.2]{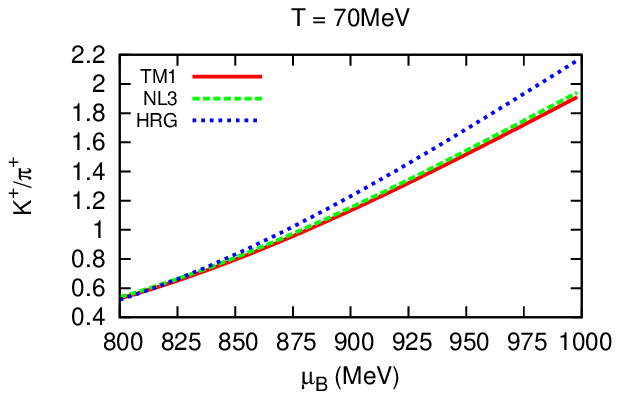}
\caption{{ $K^+/\pi^+$ as a function of baryon chemical potential $\mu_B$ at fixed values of temperature $T =50$MeV and $T= 70$MeV for (a). Strongly interacting hadronic matter with coupled baryons and bosons with two parameter sets TM1 and NL3 for kaonic potential UK[S] and UK[W] respectively, (b). For non-interacting hadrons using hadron Resonance Gas (HRG) model.}}
\end{figure}

\begin{figure}[htp]
\includegraphics[scale=1.2]{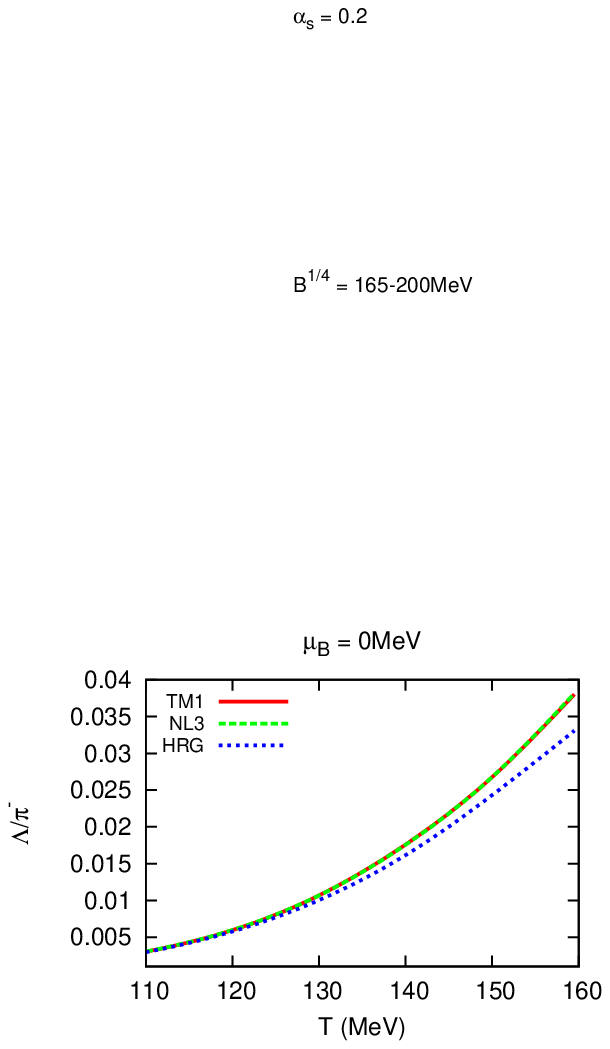}
\includegraphics[scale=1.2]{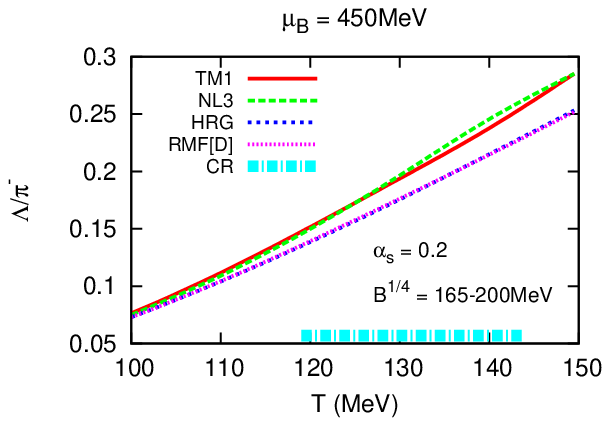}
\caption{$\Lambda/\pi^-$ as a function of temperature $T$ for fixed values of baryon chemical potential $\mu_B = 0$MeV and $\mu_B = 450$MeV for (a) Strongly interacting hadronic matter with coupled baryons and bosons with two parameter sets TM1 and NL3 for kaonic potential UK[S] and UK[W] respectively,  (b). Strongly interacting hadronic matter with decoupled baryons and bosons RMF[D] and (c). Hadron Resonance Gas (HRG) model. CR is the critical region of first order quark-hadron phase transition calculated with Bag value $B^{1/4} = 165-200$MeV. $\alpha_S$ is the coupling constant for quarks.}
\end{figure}
\begin{figure}[htp]
\includegraphics[scale=1.2]{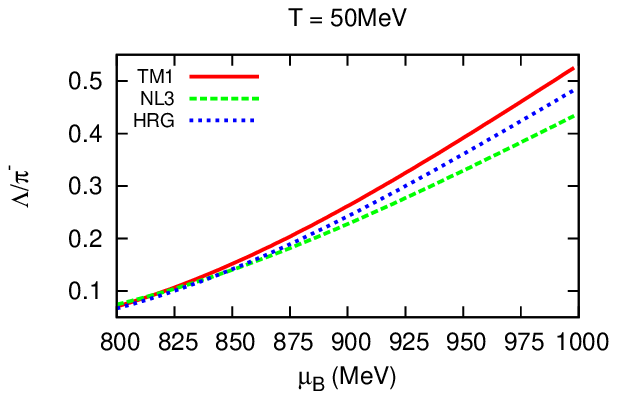}
\includegraphics[scale=1.2]{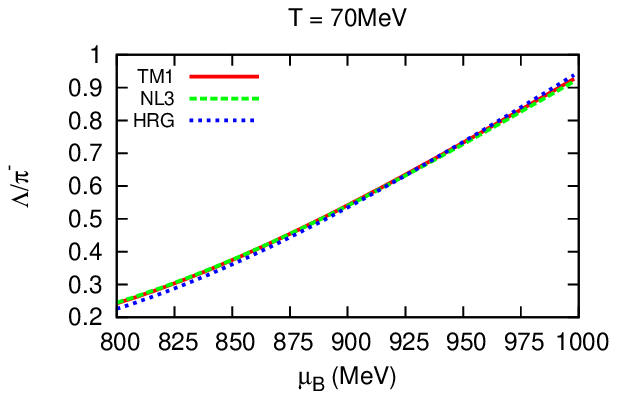}
\caption{$\Lambda/{\pi^-}$ as a function of baryon chemical potential $\mu_B$ at fixed values of temperature $T = 50$MeV and $T = 70$MeV for (a). Strongly interacting hadronic matter with coupled baryons and bosons with two parameter sets TM1 and NL3 for kaonic potential UK[S] and UK[W] respectively, (b). A system of non-interacting hadrons modelled using hadron resonance gas HRG model.}
\end{figure}

\begin{figure}[htp]
\includegraphics[scale=1.2]{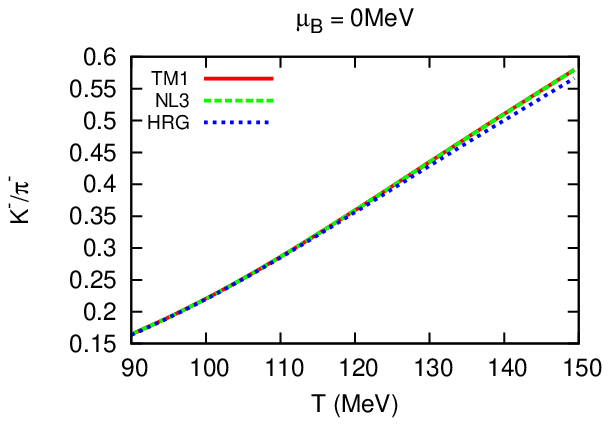}
\includegraphics[scale=1.2]{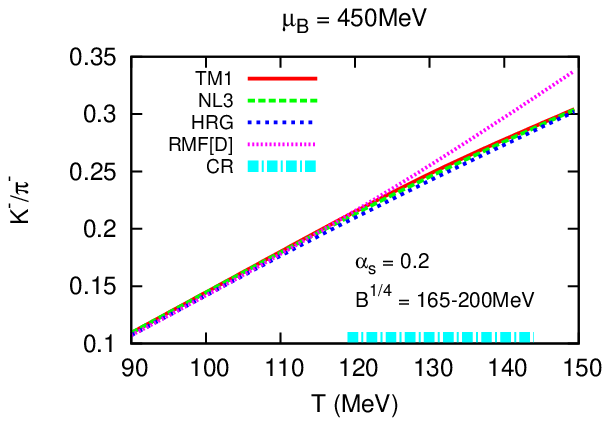}
\caption{Particle ratio $K^-/\pi^-$ as a function of temperature $T$ for different values of baryon chemical potential that is $\mu_B = 0$MeV and $\mu_B = 450$MeV for (a). Strongly interacting hadronic matter with coupled baryons and bosons with two model parameter sets TM1 and NL3 for kaonic potential UK[S] and UK[W] respectively,  (b). Strongly interacting hadronic matter with decoupled baryons and bosons RMF[D] and (c). A system of non-interacting hadrons modelled using hadron resonance gas HRG model. CR is the critical region of first order quark-hadron phase transition calculated with Bag value $B^{1/4} = 165-200$MeV. $\alpha_S$ is the strong coupling constant for quarks. }
\end{figure}
\begin{figure}[htp]
\includegraphics[scale=1.2]{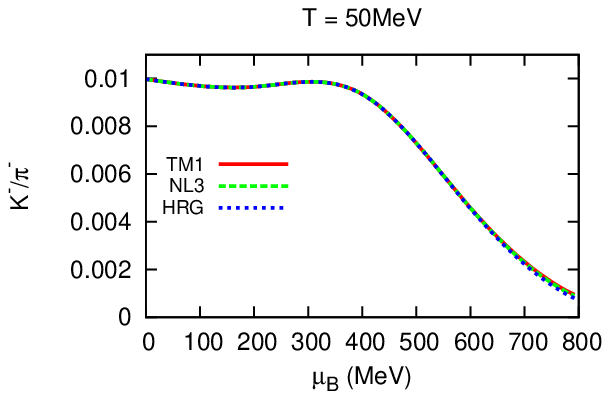}
\includegraphics[scale=1.2]{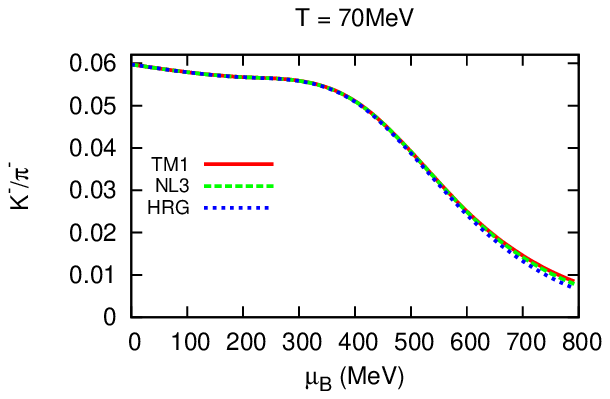}
\caption{Particle ratio $K^-/\pi^-$ as a function of baryon chemical potential $\mu_B$ for fixed values of temperature $T = 50$MeV and $T = 70$MeV for (a). Strongly interacting hadronic matter with coupled baryons and bosons with two parameter sets TM1 and NL3 for kaonic potential UK[S] and UK[W] respectively and (b). For a system with non-interacting hadrons modelled with hadron resonance gas HRG model.}
\end{figure}

\begin{figure}[htp]
\includegraphics[scale=1.2]{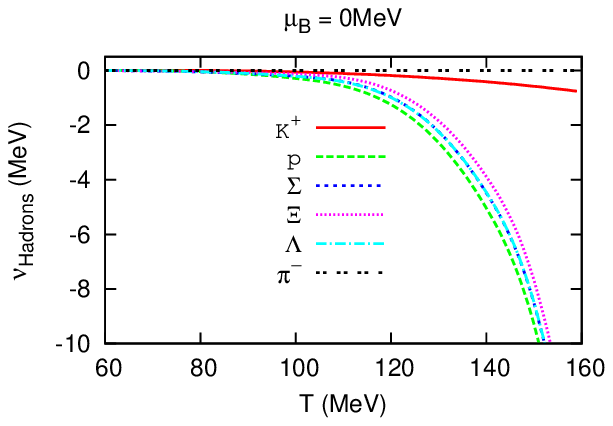}
\includegraphics[scale=1.2]{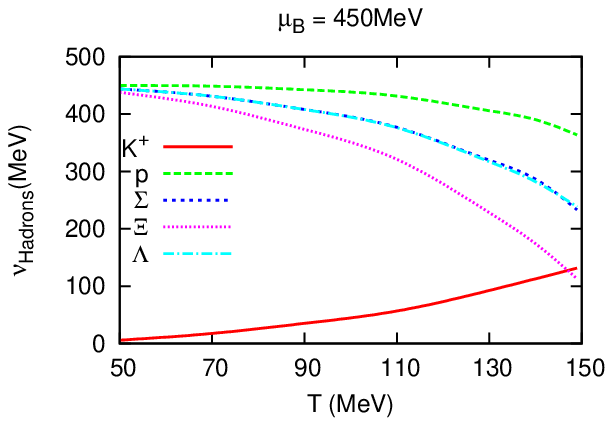}
\includegraphics[scale=1.2]{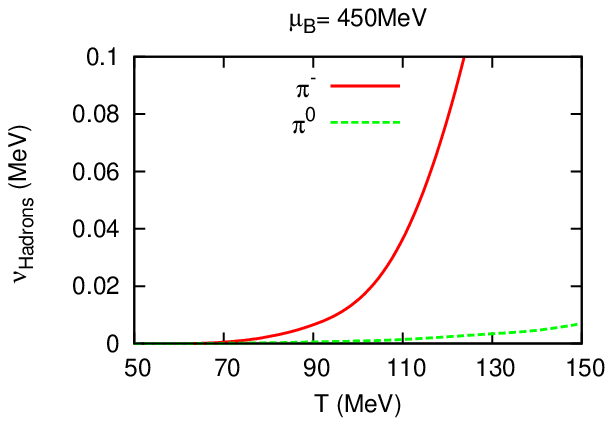}
\includegraphics[scale=1.2]{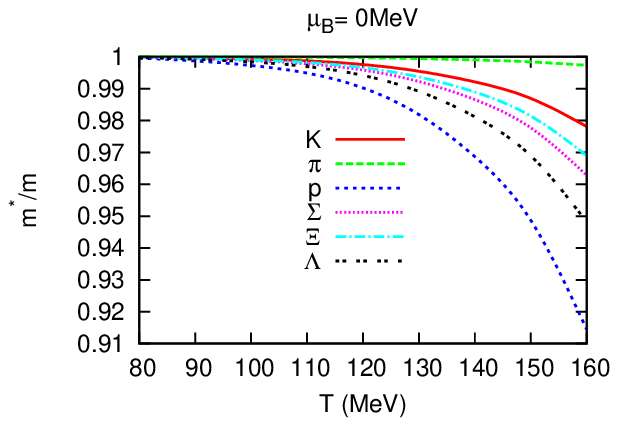}
\centering \includegraphics[scale=1.2]{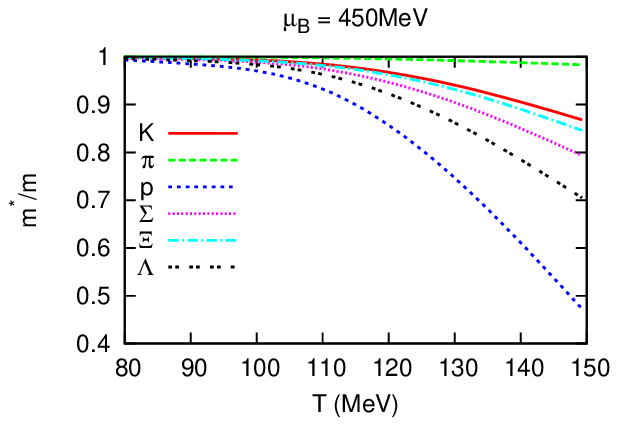}
\caption{Effective chemical potential $\nu_{Hadrons}$ and mass ratio $m^*/m$ of hadrons as a function of temperature $T$ for fixed values of baryon chemical potential $\mu_B = 0$MeV and $\mu_B = 450$MeV in case of strongly interacting hadronic matter with coupled baryons and bosons. Here parameter set TM1 is used.}
\end{figure}

\begin{figure}[htp]
\includegraphics[scale=1.2]{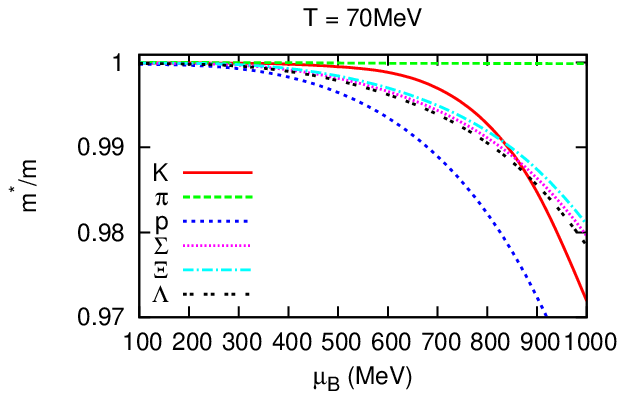}
\includegraphics[scale=1.2]{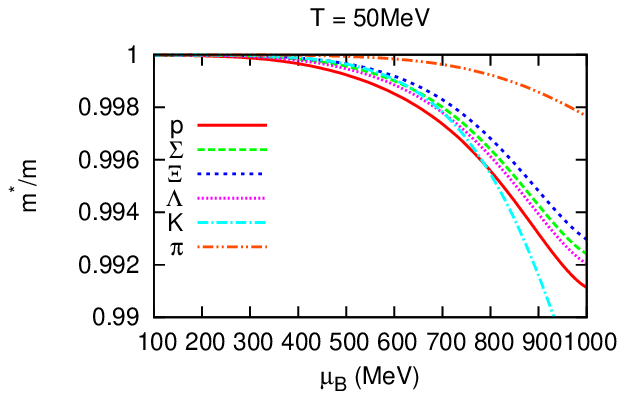}
\includegraphics[scale=1.2]{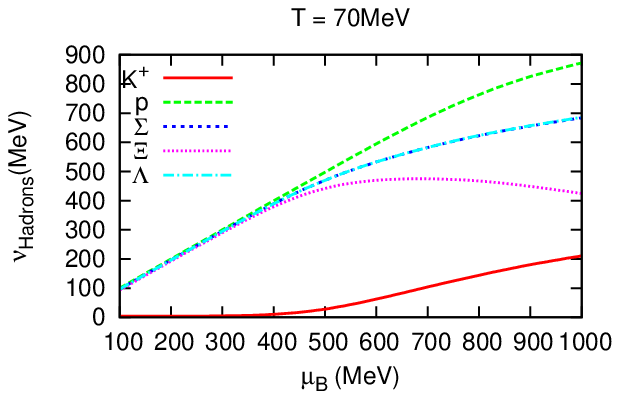}
\includegraphics[scale=1.2]{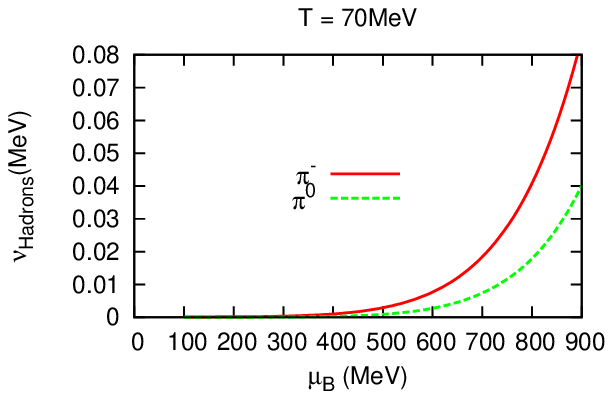}
\caption{Effective chemical potential $\nu_{Hadrons}$ and mass ratio $m^*/m$ of hadrons as a function of baryon chemical potential  $\mu_B$ for fixed values of temperature $T= 50$MeV and $T =70$MeV in case of strongly interacting hadronic matter with coupled baryons and bosons. Here parameter set TM1 is used. }
\end{figure}

\begin{figure}[htp]
\includegraphics[scale=1.2]{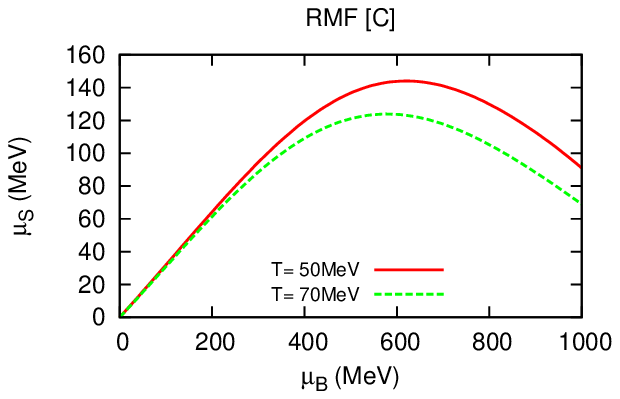}
\includegraphics[scale=1.2]{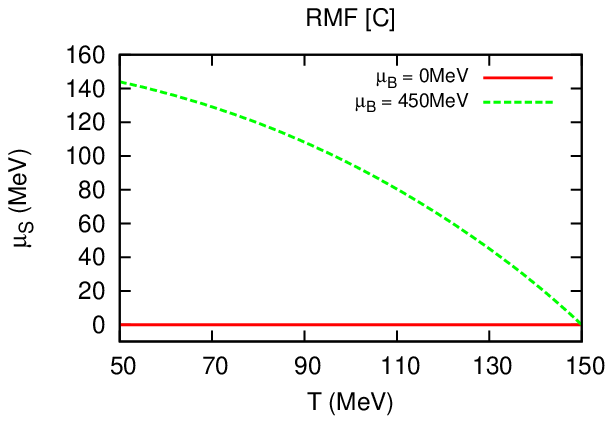}
\caption{Strange chemical potential $\mu_S$ as a function of baryon chemical potential $\mu_B$ and temperature $T$, for fixed values of temperature and baryon chemical potential, respectively, in case of a strongly interacting hadronic system with coupled baryons and bosons RMF[C]. Here parameter set TM1 is used.}
\end{figure}


\begin{thebibliography}{10}
\bibitem{wil1973}
D. J. Gross and F. Wilczek.
\newblock {\em Phys. Rev. Lett.} 30, 1343 (1973).

\bibitem{inam2016}
Inam-ul Bashir, Saeed Uddin, Hamid Nanda, arXiv:1611.04992 [nucl-th]

\bibitem{NA49}
C. Alt et al. [NA49 Collaboration]
\newblock {\em Phys. Rev. C.} 78, 034918 (2008),
J. L. Klay et al [E895 Collaboration]
\newblock {\em Phys. Rev. C.} 68, 054905 (2003),
B. I. Abelev et al. [Star Collaboration]
\newblock {\em Phys. Rev. C.} 81, 024911 (2010),
M. M. Aggarwal et al [Star Collaboration]
\newblock {\em Phys. Rev. C.} 83, 024901 (2011)



\bibitem{serot1986}
B. D. Serot and J. D. Walecka.
\newblock {\em Adv. Nucl. Phys.} 16, 1 (1986).

\bibitem{hirata1996}
P. G. Reinhard.
\newblock {\em Rept. Prog. Phys.} 52, 439 (1989).

\bibitem{Ren2002}
M. Bender, P. H. Heenen and P. G. Reinhard.
\newblock {\em Rev. Mod. Phys. }75, 121 (2003).

\bibitem{shen2006}
P. Ring.
\newblock {\em Prog. Part. Nucl. Phys.} 37, 193 (1996).



\bibitem{zim1990}
J. Meng, H. Toki, S. G. Zhou, S. Q. Zhang, W. H. Long and L. S. Geng.
\newblock {\em Prog. Part. Nucl. Phys.} 57, 470 (2006).

\bibitem{glad19855}
N. K. Glendenning.
\newblock {\em Astrophys. J.} 293, 470 (1985).


\bibitem{del1995}
Wei-Chia Chen and J. Piekarewicz.
\newblock {\em Phys. Rev. C.} 90, 044305 (2014).


\bibitem{shen1998}
H. Shen, H. Toki, K. Oyamatsu and K. Sumiyoshi.
\newblock {\em Prog. Theor. Phys.} 100, 1013 (1998).

\bibitem{shen2002}
H. Shen, H. Toki,  K. Oyamatsu and K. Sumiyoshi.
\newblock {\em APJS.} 197, 20 (2011).

\bibitem{prak1994}
V. Thorsson, M. Prakash and J. M. Lattimer
\newblock {\em Nucl. Phys. A}. 572, 693 (1994).


\bibitem{glad1999}
N. K. Glendenning and J. Schaffner-Bielich
\newblock {\em Phys. Rev. C}. 60, 025803 (1999).


\bibitem{kap1989}
D. B. Kaplan and A. E. Nelson
\newblock {\em Phys. Lett. B}. 175, 57 (1986).


\bibitem{ellis1995}
P. J. Ellis, R. Knorren and M. Prakash
\newblock {\em Phys. Lett. B}. 349, 11 (1995).

\bibitem{ellis2}
P. J. Ellis, R. Knorren and M. Prakash
\newblock {\em Phys. Rev. C}. 52, 3470 (1995).


\bibitem{sugahara1994}
Y. Sugahara and H. Toki.
\newblock {\em Nucl. Phys. A}. 579, 557 (1994).

\bibitem{ring1997}
G. A. Lalazissis, J. Konig and P. Ring.
\newblock {\em Phys. Rev. C}. 55, 540 (1997).


\bibitem{gal1988}
D. J. Millener, C. B. Dover and A. Gal.
\newblock {\em Phys. Rev. C}. 38, 2700 (1988).

\bibitem{gal2000}
J. Schaffner and A. Gal.
\newblock {\em Phys. Rev. C}. 62, 034311 (2000).

\bibitem{bunta2004}
J. K. Bunta and S. Gmuca.
\newblock {\em Phys. Rev. C}. 70, 054309 (2004).

\bibitem{schaffner1994}
J. Schaffner, C. B. Dover, A. Gal, C. Greiner, D. J. Millener and H. Stocker
\newblock {\em Ann. Phys}. 235, 35 (1994).

\bibitem{shaffner1993}
J. Schaffner, C. B. Dover, A. Gal, C. Greiner and H. Stocker.
\newblock {\em Phys. Rev. Lett.} 71, 1328 (1993).


\bibitem{mis2005}
I.N. Mishustin, L.M. Satarov, T.J. Burvenich, H. Stocker, W. Greiner
\newblock {\em Phys. Rev. C.} 71, 035201 (2005).


\bibitem{soll}
J. Sollfrank and U. Heinz, arXiv: 9505004 [nucl-th]

\bibitem{satarovkarsch2001}
L. M. Satarov, M. N. Dmitriev and I. N. Mishustin  arXiv: 0901.1430 [hep-ph], F. Karsch, E. Laermann and A. Peikert.
\newblock {\em Nucl. Phys. B} 605, 579 (2001).

\bibitem{alford}
M. Alford, M. Braby, M. Paris and S. Reddy
\newblock {\em Astrophys. J} 629, 969-978 (2005)


\bibitem{thoma2000}
K. Schertler, C. Greiner, J. Schaffner-Bielich and M. H. Thoma
\newblock {\em Nucl.Phys.A}. 667, 463-490 (2000).





\bibitem{rijken}
Th. A. Rijken,  arXiv: 9401004 [nucl-th].

\newpage

\begin{center}
{\bf Table I}.\hspace{3mm}  TMI parameter set  used in our calculation. \\
\large
\vspace{5mm}
\begin{tabular}{r r r r r r r r r r r r }
\hline
\hline
&$m_N$ & $m_\sigma$ & $m_\omega$ & $m_\rho$ & $g_{\sigma N}$ & $g_{\omega N}$ & $g_{\rho N}$ & $g_2 (fm^{-1})$& $g_3$ & $c_3$ & \\
\hline
\hline
TM1& 938.0&511.198&783.0&770.0&10.029&12.614&4.632 &-7.2323& 0.618 &71.308&\\
\hline
NL3& 939.0&508.194&782.501&763.0&10.217&12.868&4.474 &-10.431& -28.885 &-&\\
\hline
\end{tabular}\\
\vspace{5mm}
\begin{center}
{\bf Table-II}. \hspace{2mm}Meson-hyperon coupling constants.\\
\vspace{5mm}
\begin{tabular}{r r r r r r }
\hline
\hline
&$g_{\sigma \Lambda}$ & $g_{\sigma \Sigma}$ & $g_{\sigma \Xi}$ & $g_{{\sigma^*} \Lambda}$ & $g_{{\sigma^*} \Xi}$  \\
\hline
\hline
TM1&6.170& 4.472 & 3.202 & 7.018 & 12.600\\
\hline
NL3&6.269& 4.709 & 3.242 & 7.138 & 12.809\\
\hline
\end{tabular}
\end{center}
\vspace{5mm}
{\bf Table III}. \hspace{2mm}Kaon-meson coupling constants.\\
\vspace{2mm}
\begin{tabular}{r  r r r r r r r }
\hline
\hline
Set &$\hspace{3mm}$ TM2 &$\hspace{3mm}$ GL85\\
\hline
& UK[S] & UK[W]\\
\hline
Ref. & \cite{sugahara1994} & \cite{glad19855}\\

$g_{\sigma K}$&   2.27 & 1.27\\
$g_{\omega K}$&  3.02 & 3.02\\
$g_{\rho K}$&  3.02 & 3.02\\
$g_{\sigma^* K}$& 2.65 & 2.65\\
$g_{\phi K}$ & 4.27 & 4.27\\
\hline
\hline
\end{tabular}
\end{center}
\vspace{5mm}
\begin{center}
{\bf Table IV}. \hspace{2mm}Pion-meson coupling constants \cite{rijken}.\\
\vspace{2mm}
\begin{tabular}{r  r r r r r r r }
\hline
\hline
$g_{\sigma \pi}$ \hspace{5mm} & $g_{\omega \pi}$ \hspace{5mm} & $g_{\rho \pi}$ \hspace{5mm} & $g_{\sigma^* \pi}$ \hspace{5mm} & $g_{\phi \pi}$\\
\hline
-0.170 $\hspace{5mm}$&-0.001 $\hspace{5mm}$&0.506$\hspace{5mm}$ &0.0 $\hspace{5mm}$&0.0\\
\hline
\end{tabular}
\end{center}





\end{thebibliography}
\end{document}